\documentclass[conference]{IEEEtran}
\usepackage{amsmath,amssymb}
\usepackage{amsfonts}
\usepackage[dvips]{graphicx}
\usepackage{dsfont}
\usepackage{algorithmicx}
\usepackage{balance}
\usepackage{algorithm}
\usepackage{mathtools}
\usepackage{comment}

\usepackage{color}
\usepackage{pgfplots}           %for plotting graphs
\pgfplotsset{compat=1.16}
\usepackage{caption}

\usepackage{enumerate}
\usepackage{graphics}
\usepackage{cite}
\usepackage{mathrsfs}
\usepackage{subcaption}
\usepackage{stfloats}
\usepackage{xfrac}

\usepackage{tikz}
\usepackage{mathdots}
\usepackage{yhmath}
\usepackage{cancel}
\usepackage{color}
\usepackage{siunitx}
\usepackage{array}
\usepackage{multirow}
\usepackage{textcomp}
\usepackage{gensymb}
\usepackage{tabularx}
\usepackage{booktabs}
\usetikzlibrary{fadings}
\usetikzlibrary{patterns}
\usetikzlibrary{shadows.blur}
\usepackage{graphicx}
\usepackage{bm}
\usepackage{subcaption}
\usepackage{stfloats}

\usepackage{mathtools, nccmath}
\usepackage{algpseudocode}
\usepackage{pgfplots}

\begin{document}

\title{\huge Realizing RF Wavefront Copying with
RIS \\ for Future Extended Reality Applications}

\author{\IEEEauthorblockN{Stavros Tsimpoukis\IEEEauthorrefmark{1}, Dimitrios Tyrovolas\IEEEauthorrefmark{2}\IEEEauthorrefmark{3}, Sotiris Ioannidis\IEEEauthorrefmark{3}, \\ Ian F. Akyildiz\IEEEauthorrefmark{4}, George K. Karagiannidis\IEEEauthorrefmark{2}\IEEEauthorrefmark{5}, Christos Liaskos\IEEEauthorrefmark{1}\IEEEauthorrefmark{6}}

\IEEEauthorblockA{\IEEEauthorrefmark{1}Institute of Computer Science, Foundation for Research and Technology Hellas, Greece,}
\IEEEauthorblockA{e-mail: \{statsimp, cliaskos\}@ics.forth.gr}
\IEEEauthorblockA{\IEEEauthorrefmark{2}Department of Electrical and Computer Engineering, Aristotle University of Thessaloniki, 54124 Thessaloniki, Greece,}
\IEEEauthorblockA{e-mail: \{tyrovolas, geokarag\}@auth.gr}
\IEEEauthorblockA{\IEEEauthorrefmark{3} Dept. of Electrical and Computer Engineering, Technical University of Crete, Chania, Greece, e-mail: sotiris@ece.tuc.gr}
\IEEEauthorblockA{\IEEEauthorrefmark{4} Truva Inc., Alpharetta, GA 30022, USA, email: ian@truvainc.com}
\IEEEauthorblockA{\IEEEauthorrefmark{5}Artificial Intelligence \& Cyber Systems Research Center, Lebanese American University (LAU), Lebanon}
\IEEEauthorblockA{\IEEEauthorrefmark{6}Computer Science Engineering Department, University of Ioannina, Ioannina, Greece.} 
\vspace{-6mm}
}

\maketitle	

\begin{abstract}

Lately a new approach to Extended Reality (XR), denoted as XR-RF, has been proposed which is realized by combining Radio Frequency (RF) Imaging and programmable wireless environments (PWEs). RF Imaging is a technique that aims to detect geometric and material features of an object through RF waves. On the other hand, the PWE focuses on the the conversion of the wireless RF propagation in a controllable, by software, entity through the utilization of Reconfigurable Intelligent Surfaces (RISs), which can have a controllable interaction with impinging RF waves. In that sense, this dynamic synergy leverages the potential of RF Imaging to detect the structure of an object through RF wavefronts and the PWE's ability to selectively replicate those RF wavefronts from one spatial location to wherever an XR-RF mobile user is presently located. Then the captured wavefront, through appropriate hardware, is mapped to the visual representation of the object through machine learning models. As a key aspect of the XR-RF's system workflow is the wavefront copying mechanism, this work introduces a new PWE configuration algorithm for XR-RF. Moreover, it is shown that the waveform replication process inevitably yields imprecision in the replication process. After statistical analysis, based on simulation results, it is shown that this imprecision can be effectively modeled by the gamma distribution.

\end{abstract}
%\vspace{-0.2cm}

%\vspace{-0.3cm}
\section{Introduction}\label{S:Intro}

Recently, an innovative approach has been introduced, wherein wireless propagation is redefined as a software-defined phenomenon \cite{liaskos2018new}. This concept referred to as Programmable Wireless Environments (PWEs) has demonstrated considerable promise in enhancing wireless communications. Furthermore, extending beyond the domain of communications, this paradigm has illustrated the capability to enhance RF imaging, facilitating the direct generation of graphics that can compete with existing extended reality (XR) technology. This emergent synergy of RF imaging and PWEs is denoted as XR-RF~\cite{liaskos2022xr}. 

In order to realize the paradigm of PWE the employment of the  Reconfigurable Intelligent Surfaces (RISs) technology is a key-enabler factor. Specifically, RISs are specially engineered planar structures composed of RF resonating elements that can dynamically manipulate the characteristics of the impinging upon them RF waves, including their direction, intensity, phase shift and polarization \cite{zeng2021reconfigurable, cscn2022}. Therefore, by coating large surfaces (e.g., walls) in an environment with RISs, the PWE is created where the wireless propagation can be transformed into a controllable process, and thus new applications can emerge. Such an application is the aforementioned XR-RF concept, in which 
an "RF-illuminated" object scatters an RF wave in a distinctive manner due to its unique geometry. The scattered RF wavefront, which incorporates the RF information that describes the object, is then directed towards the user through the RISs, and is finally translated into a visual representation via AI tools~\cite{liaskos2022xr}. Critical to the XR-RF's user experience is the ability of the PWE to adapt to user's movement through space. As the users move, the XR-RF service must adapt to their perspective, ensuring that the displayed objects align with their changing viewpoints. Addressing this crucial aspect, there emerges a need for an RF wavefront copy-paste mechanism.

\color{black}
Despite XR-RF's potential, there is limited research on RIS capabilities in replicating wavefronts. In more detail, while the authors in \cite{hu2020reconfigurable} and \cite{zhang2022toward} examine the performance boost achieved by a single RIS unit in a posture recognition application, they do not satisfy the requirements of XR-RF, as they do not devise a methodology for wavefront copying. Moreover, \cite{liu2021learning} investigated how RISs enhance XR service reliability through improved communication channels but did not explore RIS's full capabilities for novel XR methods like XR-RF, which directly manipulates wavefronts for XR. Finally, the authors in \cite{cscn2023} examined the impact of the RIS configuration on a 3D object classification task within an XR-RF scenario. To this end, to the best of the authors' knowledge, there exists no work that develops a "Copy-Paste" methodology for RF wavefronts which is essential for realizing the XR-RF concept within PWEs. 

In this paper, we present a novel methodology for RF wavefront copying within a PWE, utilizing the available RIS units to replicate the desired RF wavefront at the receiver's antennas. Specifically, we detail a system model within a two-room PWE setup, to recreate the exact RF wavefront at the receiver's location, overcoming the absence of line of sight (LoS) between the transmitter and the receiver. Furthermore, acknowledging the discrete operational modes of RISs due to the available functionalities codebook, we develop a routing algorithm that minimizes replication errors under practical PWE constraints. Moreover, it is shown that the spatial characteristics of an environment can lead to discrepancies between the original and the replicated wavefronts. Specifically, the location of RIS units in a space can affect the intended direction of arrival of rays across the replicated wavefront. Thus, the paper studies this effect and contributes a statistical model for its quantification. The study is concluded by systematically evaluating the impact of PWE characteristics, such as RIS size and the number of receiver antennas, on wavefront replication accuracy, supported by appropriate simulation results.

%In this paper, we present a novel methodology for RF wavefront copying within a PWE, utilizing the available RIS units to accurately replicate the desired RF wavefront at the receiver's antennas for precise 3D object visualization. Specifically, we detail a system model within a two-room PWE setup, to recreate the exact RF wavefront at the receiver's location, overcoming the absence of line of sight (LoS) between the transmitter and the receiver. Furthermore, acknowledging the discrete operational modes of RISs due to the available functionalities codebook, we develop a routing algorithm that minimizes replication errors under practical PWE constraints. Finally, through comprehensive noise modeling, we systematically evaluate the impact of PWE characteristics, such as RIS size and the number of receiver antennas, on wavefront replication accuracy, supported by appropriate simulation results.

The remainder of this paper is organized as follows. The system model is described in Section \ref{sysmodl}. The performance analysis of the considered network is presented in Section \ref{analysis} and the numerical results are presented in Section \ref{secnum}. Finally, Section \ref{conclusion} concludes the paper.

\section{Preliminaries \& System Model}\label{sysmodl}

\subsection{Preliminaries} 

In the context of PWEs applied to XR, the XR-RF concept emerges as a key innovation, merging RF-imaging technologies with the dynamic capabilities of PWEs. Specifically, the XR-RF takes place in an indoor scenario where a transmitter antenna targets a 3D object \begin{comment}
    across multiple rooms
\end{comment}
within a PWE setup. This system utilizes an array of RIS to facilitate the XR-RF mechanism, where users, through headsets equipped with MIMO antennas and appropriate AI software, receive RF wavefronts that are transformed into graphical outputs. The essence of XR-RF lies in capturing an RF footprint that accurately embodies the object's geometry, aiming to replicate this footprint for a user positioned elsewhere within the PWE environment. This approach ensures the geometrical details of the object are precisely captured and conveyed through the RF wavefronts directed towards a receiver antenna, which are then translated into visual graphics by neural networks.

While existing studies have examined various PWE configurations to optimize object detection in XR-RF, they reveal a significant research gap in the "copy-paste" mechanism of wavefront replication necessary for XR-RF. This challenge, centered on replicating the desired wavefront at any user location within the PWE, is affected by the complex optimization required for RIS to perform multiple RF functionalities, essential for maintaining the low latency critical to XR-RF experiences. In this direction, a predefined set of RIS configurations called codebook, is typically employed, aiming to balance efficiency with the constraints of practical implementation. However, this approach, while mitigating time complexity issues, introduces variability in the received RF wavefront, potentially compromising XR-RF experience. Consequently, there is a clear need for a method that not only enables efficient RF wavefront copying within a PWE but also aligns with the practical constraints, ensuring that users experience consistent and immersive visuals irrespective of their location within the PWE. Developing such a system is crucial for advancing XR-RF services, as it directly addresses the unexplored challenge of achieving precise wavefront replication within the operational realities of PWEs, thereby enhancing the immersive quality and accessibility of XR technologies.

\subsection{System Model}

\begin{figure*}[!t]
\centering{}\textcolor{black}{\includegraphics[clip,width=0.78\textwidth]{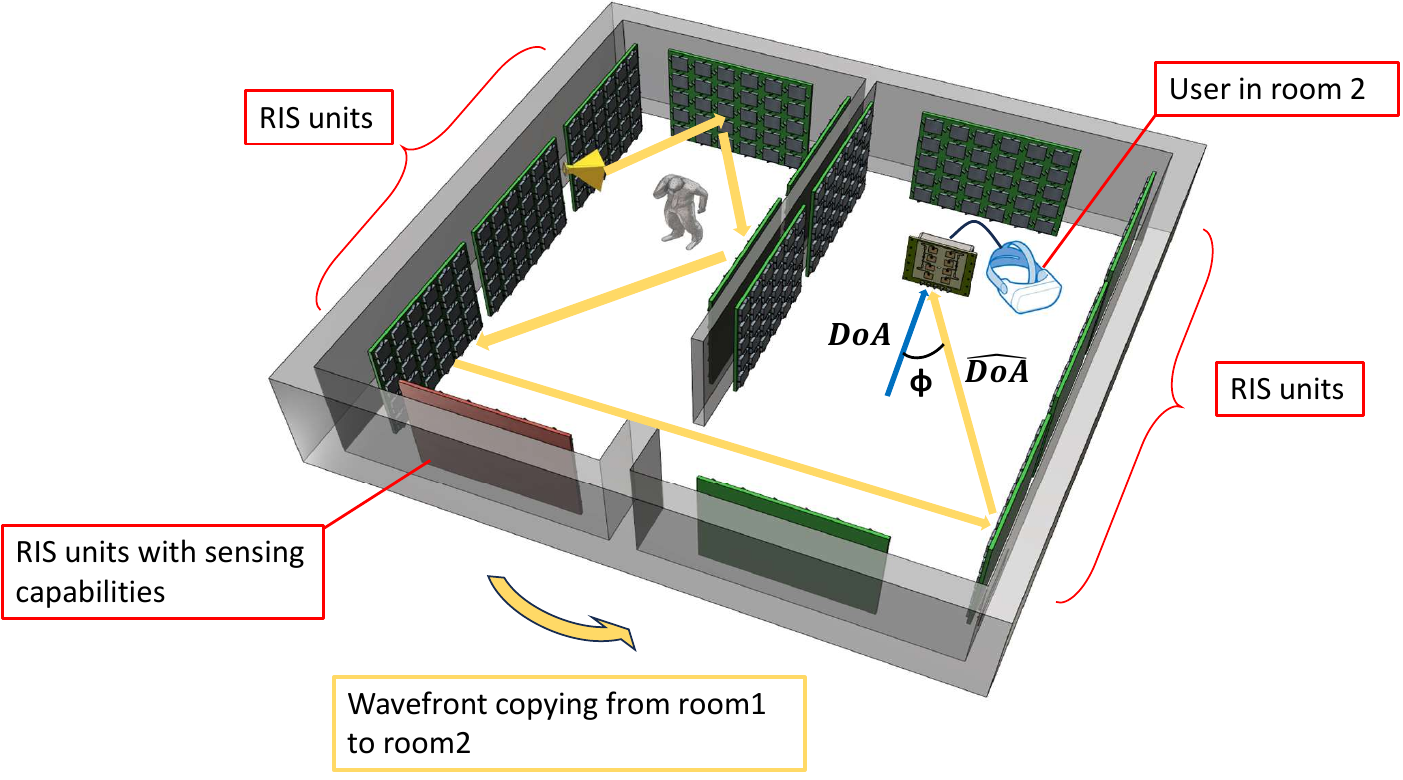}}
\caption{\protect\label{fig:System Model} Overview of the RF wavefront replication process within a two-room PWE.}
\end{figure*}

Fig. \ref{fig:System Model} illustrates a two-room PWE, where the first room contains a transmitter, an object, and an RIS equipped with sensing capabilities \cite{liaskos2019absense}, while the second room contains an XR-RF user who will receive a specifically tailored RF wavefront corresponding to the object from the first room. Moreover, the RIS units within this PWE are capable of executing a set of RF functionalities according to the assumed codebook, which includes i) \textbf{Diffusion}: the RISs are set in a random configuration to reflect impinging waves in various directions, ii) \textbf{Beam steering}: enabling the RISs to direct the incident wave towards another RIS or one of the receiver's antennas, necessitating a line of sight (LoS) connection with that RIS or antenna, and iii) \textbf{Absorption}: where the RISs are tuned to absorb the RF waves that impinge upon them.

Within the examined PWE, the XR-RF service is realized through a carefully orchestrated two-phase process. Initially, the system is activated by the transmission of a single tone from the transmitter, while setting the RIS units in the first room to adopt diffusion mode, to generate a wide range of reflections from the object, thereby producing a unique wavefront that is captured by the sensing RIS and accurately represents the object's geometrical details \cite{cscn2023}. Afterwards, the XR-RF procedure advances to the second phase, where the transmitter emits another single tone and the RISs aim to recreate the wavefront precisely towards the XR-RF user via their beam steering functionality. Finally, it should be mentioned, that the RIS units that do not contribute to the wavefront routing procedure are configured to absorption mode, to eliminate unnecessary RF waves from the transmission path and ensure reliable RF wavefront copying to the XR-RF user.

\section{Wavefront Copying}\label{analysis}

This section introduces our approach to RF wavefront copying within a PWE, focusing on the precise manipulation of RF wavefronts using RIS. Our method optimizes the replication process by ensuring minimal Direction of Arrival (DoA) deviations and employing the fewest possible RIS units for directing the wavefront. In the following subsections, we present the routing algorithm that minimizes DoA deviations and the number of utilized RIS in the wavefront directing process. Additionally, we perform statistical modeling of the deviations between the received wavefront and the desired one to deduce the effect of the PWE characteristics on the wavefront copying performance. 

\subsection{Wavefront Routing Methodology}
A PWE can effectively be modeled as a graph $G(V,E)$, offering a structured approach to configuring its components for optimized XR-RF processes \cite{liaskos2019network}. In this graph model, each element of the PWE, including XR-RF devices, transmitter, and RIS units, is represented as vertices ($v \in V$). Furthermore, an edge ($e \in E$) connects two vertices if there is a direct LoS link between them, facilitating a clear depiction of potential communication paths within the PWE. In this direction, by defining the subsets $E_u \subset E$ and $E_t \subset E$ that outline the connections from users to RIS units and from the transmitter to RIS units, we can address the wavefront copying challenge as a path-finding problem, while taking into account the PWE characteristics.

\begin{algorithm}
    \caption{Wavefront Routing Algorithm}
    \begin{algorithmic}[1]
    \Procedure{getRoutes}{$RISs$, $DoAs$, $ant$}
        \State $routes \gets \mathrm{null}$
        \For{$i$ in $length(ant)$}
            \For{$j$ in $length(walls)$}
            \State $ d \gets \frac { \left( \bm{p_{0j}} - \bm{ant}(i) \right)  \cdot \bm{n_{j}}}      { \bm{DoAs}(i) \cdot \bm{n_{j}}}$
            \State $p \gets \bm{ant}(i) + d \cdot\bm{DoAs}(i)$
            \If{$p \in walls$}
                \State $point \gets p$
                \State $break$
            \Else
                \State $continue$
            \EndIf
            \EndFor
            
            \State $lastRIS \gets \arg \min_{r\in RISs}\hspace{4px}\lvert\lvert r-point\rvert\rvert$ 
            \State REMOVE($RISs$, $lastRIS$)
            \State $path \gets breadth\_first\_search(lastRIS)$
           \State APPEND($routes$, $path$)
        \EndFor
        \State \textbf{return} $routes$ 
    \EndProcedure
    \end{algorithmic}
    \label{alg:getLastTile}
\end{algorithm}

The establishment of the path-finding framework in a PWE is inherently dependent on the precise identification of the last RIS units which will steer the RF waves, ensuring their successful arrival at receiver antennas with specific Directions of Arrival (DoAs) that correspond to the desired RF wavefront. This necessity arises from the steering limitations imposed by the RIS units' codebook capabilities, which constrict the RF wavefront's directionality, making the acquisition of the RIS units that minimize DoA deviation a pivotal task. To address this challenge, we introduce the "getRoutes" algorithm, designed to select RIS units that can direct the wavefront towards the receiver antennas with the least deviation from the desired DoAs. Initially, the algorithm assigns to each receiver antenna a distinct DoA that needs replication, which is achieved by defining "$DoAs$" as the target unit vectors denoting the exact DoAs expected at the receiver antennas. Furthermore, to detect the RIS units that will be utilized as the last RISs in the routing process for each of the receiver's antennas, i.e., $lastRISs$, we identify a point $\bm{p}$ belonging to one of the indoor environment's walls, which is determined by the intersection of the wall planes with the line equation determined by the DoA unit vector and the antenna position, and is given as
\begin{equation}
    \bm{p} = \bm{ant}\left(i\right) + d\cdot\bm{DoAs}\left(i\right),
\end{equation}
where $\bm{ant}$ is an array containing the positions of the receiver antennas, $\bm{DoAs}$ is an array containing the desired DoA direction vectors for each antenna, and $d$ is the scaling factor of the line equation given by
\begin{equation}
    d = \frac{\left(\bm{p_{0j}-\bm{ant}(i)}\right) \cdot \bm{n_j}}{\bm{DoAs}(i)\cdot \bm{n_j}},
\end{equation}
where $\bm{p_{0j}}$ and $\bm{n_j}$ are the describing parameters of each wall plane equation defined as $\left(\bm{p}-\bm{p_{0j}}\right)\cdot \bm{n_j}$. Thus, by utilizing $\bm{p}$ and the positions of the RIS centers with which the receiver antennas share a LoS connection, we can select the RISs with the least Euclidean distance to $\bm{p}$, which corresponds to the RIS that will invoke the smallest DoA deviation. Finally, after determining the $lastRISs$, they are used as reference vertices for the path-finding challenge, which is approached through a breadth-first search to determine the shortest paths that also reduce the number of RIS units involved, hence, presenting an appropriate strategy for enhanced wavefront replication in an XR-RF setup.

\subsection{Wavefront Deviation Modeling}

Achieving perfect wavefront routing within a PWE is challenged by deviations from the intended DoAs, primarily due to the predefined capabilities of RIS described by the available codebook. Specifically, as it can be seen in Fig. \ref{fig:System Model}, these deviations arise since the available RIS units cannot steer impinging waves towards the receiver antennas perfectly due to their position and the codebook guidelines. Furthermore, when the optimal RIS unit for steering the wavefront towards a specific receiver antenna is already engaged, alternative RIS units must be employed, which may not offer the same level of directional precision. To that end, it becomes imperative to statistically model the wavefront deviations, to effectively characterize the impact of PWE characteristics on the DoA deviations. In this direction, based on various PWE parameters such as the number of receiver antennas and the RIS sizes, we create datasets containing the angle differences between the desired and the actual DoA, i.e., $\hat{DoAs}$, allowing the analysis of data histograms. In more detail, the aforementioned angle difference can be modeled through an appropriate distribution that best captures the observed data form. Therefore, based on the data patterns evident in the histograms in Fig.\ref{fig:2}, we select to examine the Gamma and Rayleigh distributions whose probability density functions (PDFs) are given respectively as
\begin{equation}
    f(x; k, \theta) = \frac{1}{\Gamma(k)\theta^{k}} x^{k-1}e^{-\frac{x}{\theta}}, \; \;\forall x \geq 0,
\end{equation}
where $\Gamma\left(\cdot\right)$ is the gamma function, while $k$ and $\theta$ correspond to the Gamma distribution scale and shape parameters, respectively, and
\begin{equation}
    f(x; \sigma) = \frac{x}{\sigma^2} e^{-\frac{x^2}{2\sigma^2}}, \; \; \forall x \geq 0,
\end{equation}
where $\sigma$ corresponds to the scale parameter of Rayleigh distribution. To be more precise, the parameter $k$ of the Gamma distribution can describe the frequency of smaller versus larger deviations, while $\theta$ can indicate the spread of deviations. Similarly, the Rayleigh distribution, through its scale parameter $\sigma$, can quantify the extent of DoA deviation from the targeted DoA, where an increase in $\sigma$ denotes a greater average deviation, thereby offering a comprehensive view of $\phi$. Thus, the Gamma and Rayleigh distributions allow for a clear understanding of $\phi$ while offering a solid foundation for assessing the effect of the PWE characteristics on the wavefront copying accuracy.

To acquire the parameters that best fit the data for the selected distributions, we can utilize the Maximum Likelihood Estimation technique in which we calculate the parameter vector $\bm{\lambda}$ that maximizes the function $ \ln(L(\bm{\lambda}))$, where $L(\bm{\lambda})$ is equal to
\begin{equation}
    L(\bm{\lambda}) = \prod_{i=1}^{N}f(x_i; \bm{\lambda}),
\end{equation}
where $x_i$ is the $i$-th sample of the dataset $X_d$, and $N$ equals to the number of the dataset samples. Thus, considering that the parameter vector $\bm{\lambda} = \left[ k, \theta\right]$ for the Gamma distribution, we can obtain $\hat{k}$ and $\hat{\theta}$, which are the estimate values of $k$ and $\theta$, respectively, that optimize the data fitting procedure. Specifically, $\hat{k}$ can be calculated numerically as the root of $g(\cdot)$, which is given as
\begin{equation} \label{eq:kappa}
    g(\hat{k}) = 
    \ln{(\hat{k})} - \ln{\psi(\hat{k})} - \ln{(\overline{X_d})} + \overline{\ln{(X_d)}},
\end{equation}
where $\psi(\cdot)$ is the digamma function, $\overline{X_d}$ denotes the mean value of $X_d$ values, and $\overline{\ln{(X_d)}}$ is the mean value of the dataset $\ln(X_d)$, whereas $\hat{\theta}$ can be calculated as
\begin{equation}\label{eq:theta}
    \hat{\theta}=\frac{\overline{X_d}}{\hat{k}}.
\end{equation}
Similarly for the Rayleigh distribution, considering that $\bm{\lambda} = \sigma$, we can calculate $\hat{\sigma}$ which is given as
    \begin{equation}\label{eq:sigma}
        \hat{\sigma} = \sqrt{\frac{1}{2N} \sum_{i=1}^{N} {x^2_{i}}}.
    \end{equation}
Finally, to evaluate the accuracy of the derived deviation models, we can compare the relative entropy of the original data and random samples generated from the derived distributions through the Kullback-Leibler Divergence (KLD) which is a measure of how much the random samples are different from the original ones, and is expressed as
\begin{equation}\label{eq:KLD}
    D_{KL}(P||Q) = \int_{-\infty}^{\infty}p(x)\log{\frac{p(x)}{q(x)}}dx,
\end{equation}
where $p(x)$ is the empirical distribution of the original data and $q(x)$ is the distribution of the random samples.

\begin{figure}[h!]
    \centering
    \begin{minipage}{.5\textwidth}
        \centering
\begin{tikzpicture}
        \begin{axis}[
	width=0.95\linewidth,
	xlabel = {$\bm{\phi}$},
	ylabel = {},
	ymin = 0,
	ymax = 0.22,
        yticklabel style={
        /pgf/number format/fixed,
        /pgf/number format/precision=5
    },
    xticklabel={$\pgfmathprintnumber{\tick}^{\circ}$},
    scaled y ticks=false,
        area style,
    ]
    \addplot+[ybar interval,mark=no] plot coordinates { (0, 0.086) (1.4, 0.2) (2.8, 0.17) (4.2, 0.1) (5.6, 0.065) (7, 0.041) (8.4, 0.022) (9.8, 0.013) (11.2, 0.0075) (12.6, 0.002)}; \addlegendentry{$M=4\times4$}

    \addplot+[ybar interval,mark=no, opacity=0.67] plot coordinates { (0, 0.087) (1.5, 0.2) (3.0, 0.147) (4.5, 0.088) (6.0, 0.056) (7.5, 0.036) (9, 0.017) (10.5, 0.012) (12.0, 0.011) (13.5, 0.0064)}; \addlegendentry{$M=6\times6$}

    \addplot+[ybar interval,mark=no, opacity = 0.63] plot coordinates { (0, 0.103) (1.8, 0.189) (3.6, 0.107) (5.4, 0.062) (7.2, 0.036) (9.0, 0.024) (10.8, 0.014) (12.6, 0.0096) (14.4,  0.0053) (16.2, 0.0037)}; \addlegendentry{$M=8\times8$}

    \addplot+[ybar interval,mark=no, opacity=0.57] plot coordinates { (0, 0.122) (2.4, 0.142) (4.8, 0.063) (7.2, 0.0332) (9.6, 0.0208) (12, 0.0118) (14.4, 0.0071) (16.8, 0.0068) (19.2, 0.0051) (21.6, 0.0036)}; \addlegendentry{$M=10\times10$}
        
        \end{axis}
	\end{tikzpicture}
        \label{fig:2a}
    \end{minipage}%

\begin{minipage}{.5075\textwidth}
        \centering
\begin{tikzpicture}
        \begin{axis}[
	width=0.95\linewidth,
	xlabel = {$\bm{\phi}$},
	ylabel = {},
	ymin = 0,
	ymax = 0.55,
        area style,
        xticklabel={$\pgfmathprintnumber{\tick}^{\circ}$},
    ]
    \addplot+[ybar interval,mark=no] plot coordinates { (0, 0.2) (0.45, 0.524) (0.9, 0.539) (1.35, 0.343) (1.8, 0.233) (2.25, 0.165) (2.7, 0.09) (3.15, 0.057) (3.6, 0.046) (4.05, 0.021)}; \addlegendentry{$d_r=0.15$m}
    
    \addplot+[ybar interval,mark=no, opacity=0.66] plot coordinates { (0, 0.123) (0.77, 0.317) (1.54, 0.316) (2.31, 0.197) (3.08, 0.129) (3.85, 0.096) (4.62, 0.049) (5.39, 0.031) (6.16, 0.025) (6.93, 0.012)}; \addlegendentry{$d_r=0.25$m}
    
    \addplot+[ybar interval,mark=no, opacity=0.62] plot coordinates { (0, 0.08) (1.1, 0.238) (2.2, 0.216) (3.3, 0.141) (4.4, 0.093) (5.5, 0.06) (6.6, 0.041) (7.7, 0.015) (8.8, 0.015) (9.9, 0.005)}; \addlegendentry{$d_r=0.35$m}
    
    \addplot+[ybar interval,mark=no, opacity=0.57] plot coordinates { (0, 0.071) (1.6, 0.174) (3.2, 0.142) (4.8, 0.096) (6.4, 0.056) (8, 0.036) (9.6, 0.023) (11.2, 0.011) (12.8, 0.009) (14.4, 0.002)}; \addlegendentry{$d_r=0.45$m}
        \end{axis}
	\end{tikzpicture}
        \label{fig:2b}
    \end{minipage}%
    \caption{Deviations of $\phi$ for a) $d_r=0.40$m and various $M$ values b) $M=4\times4$ and various $d_r$ values}
    \label{fig:2}
\end{figure}

\section{Simulation Results}\label{secnum}
In this section, we present simulation results for the scenario illustrated in Fig. \ref{fig:System Model} where a single-antenna transmitter is located in the left room of a PWE consisting of RISs with dimensions equal to $d_r\times d_r$ where $d_r$ equals to the length of each RIS side, while an XR-RF user with $M$ antennas is located within the right room. Specifically, given $d_r$ and $M$, by applying the proposed wavefront routing methodology, we obtain the angle difference for each of the receiver antennas between $\hat{DoAs}$ and the $DoAs$ for 100 different wavefront cases.

%In our study, 100 different wavefront copying scenarios were simulated for each PWE parameter combination (e.g. 100 examples for Rec. Antennas= 10x10 and RIS size = 0.15m etc.). The system model parameters that were considered for the Rx Dimensions were: \{4x4, 6x6, 8x8, 10x10\}. For the RIS sizes that depends on the Rx dimension. 
%For the case of:
%\begin{itemize}
%    \item 
%    4x4: The RIS sizes examined are\{0.15, 0.20, 0.25, ... , 1\}m

%    \item 
%    6x6: The RIS sizes examined are \{0.15, 0.20, 0.25, ... , 0.75\}m
%    \item 
%    8x8: The RIS sizes examined are \{0.15, 0.20, 0.25, ... , 0.55\}m
%    \item 
%    10x10: The RIS sizes examined are \{0.15, 0.20, 0.25, ... , 0.55\}m
%\end{itemize}

%The reason that there is a difference in the experiments, concerning the sizes of the RIS units, is that as the receiver array dimension increases so do the antennas that need to be serviced. After a certain threshold value, for each Rx dimension, the $getLastRIS$ algorithm can not be implemented because there are not enough RISs to service the receiver antennas. 

\begin{figure}[h!]
    \centering
    \begin{minipage}{.5\textwidth}
        \centering
\begin{tikzpicture}
        \begin{axis}[
	width=0.95\linewidth,
	xlabel = {$d_r$ (m)},
	ylabel = {$\hat{k}$},
	xmin = 0.15,
        xmax = 0.55,
	ymin = 1.2,
	ymax = 2.6,
	xtick = {0.15,0.2,...,0.55},
	grid = major,
        legend style = {font = \scriptsize},
	legend cell align = {left},
	legend pos = south west,
	]
        \addplot[
	black,
	mark=square*,
        mark repeat=1,
	mark size = 2,
 	line width = 1pt,
	style = solid,
	]
	table{data/kappa4.dat};
	\addlegendentry{$M= 4\times4$}
         \addplot[
	black,
	mark=triangle*,
        mark repeat=1,
	mark size = 3,
 	line width = 1pt,
	style = solid,
	]
	table{data/kappa6.dat};
	\addlegendentry{$M= 6\times6$}
         \addplot[
	black,
	mark=diamond*,
        mark repeat=1,
	mark size = 3,
 	line width = 1pt,
	style = solid,
	]
	table{data/kappa8.dat};
	\addlegendentry{$M= 8\times8$}
         \addplot[
	black,
	mark=*,
        mark repeat=1,
	mark size = 2,
 	line width = 1pt,
	style = solid,
	]
	table{data/kappa10.dat};
	\addlegendentry{$M= 10\times10$}
        \end{axis}
	\end{tikzpicture}
        %\subcaption{$M=64$}
        \label{fig:4a}
    \end{minipage}%

\begin{minipage}{.5\textwidth}
        \centering
\begin{tikzpicture}
        \begin{axis}[
	width=0.95\linewidth,
	xlabel = {$d_r$ (m)},
	ylabel = {$\hat{\theta}$},
	xmin = 0.15,
        xmax = 0.55,
	ymin = 0.5,
	ymax = 12.0,
	xtick = {0.15,0.2,...,0.55},
	grid = major,
        legend style = {font = \scriptsize},
	legend cell align = {left},
	legend pos = north west,
	]
        \addplot[
	black,
	mark=square*,
        mark repeat=1,
	mark size = 2,
 	line width = 1pt,
	style = solid,
	]
	table{data/theta4.dat};
	\addlegendentry{$M= 4\times4$}
         \addplot[
	black,
	mark=triangle*,
        mark repeat=1,
	mark size = 3,
 	line width = 1pt,
	style = solid,
	]
	table{data/theta6.dat};
	\addlegendentry{$M= 6\times6$}
         \addplot[
	black,
	mark=diamond*,
        mark repeat=1,
	mark size = 3,
 	line width = 1pt,
	style = solid,
	]
	table{data/theta8.dat};
	\addlegendentry{$M= 8\times8$}
         \addplot[
	black,
	mark=*,
        mark repeat=1,
	mark size = 2,
 	line width = 1pt,
	style = solid,
	]
	table{data/theta10.dat};
	\addlegendentry{$M= 10\times10$}
        \end{axis}
	\end{tikzpicture}
        %\subcaption{$M=1024$}
        \label{fig:M1024}    %\subcaption{$M=1024$}
    \end{minipage}

\begin{minipage}{.5\textwidth}
        \centering
\begin{tikzpicture}
        \begin{axis}[
	width=0.95\linewidth,
	xlabel = {$d_r$ (m)},
	ylabel = {$\hat{\sigma}$},
	xmin = 0.15,
        xmax = 0.55,
	ymin = 1.2,
	ymax = 13.0,
	xtick = {0.15,0.2,...,0.55},
	grid = major,
        legend style = {font = \scriptsize},
	legend cell align = {left},
	legend pos = north west,
	]
        \addplot[
	black,
	mark=square*,
        mark repeat=1,
	mark size = 2,
 	line width = 1pt,
	style = solid,
	]
	table{data/sigma4.dat};
	\addlegendentry{$M= 4\times4$}
         \addplot[
	black,
	mark=triangle*,
        mark repeat=1,
	mark size = 3,
 	line width = 1pt,
	style = solid,
	]
	table{data/sigma6.dat};
	\addlegendentry{$M= 6\times6$}
         \addplot[
	black,
	mark=diamond*,
        mark repeat=1,
	mark size = 3,
 	line width = 1pt,
	style = solid,
	]
	table{data/sigma8.dat};
	\addlegendentry{$M= 8\times8$}
         \addplot[
	black,
	mark=*,
        mark repeat=1,
	mark size = 2,
 	line width = 1pt,
	style = solid,
	]
	table{data/sigma10.dat};
	\addlegendentry{$M= 10\times10$}
        \end{axis}
	\end{tikzpicture}
    \end{minipage}
    \caption{Estimated Distribution Parameters versus $d_r$ for different $M$ values}
    \label{fig:3}
\end{figure}

In Fig. \ref{fig:3}, we illustrate the effect of $d_r$ on the values of the estimated parameters $\hat{k}$, $\hat{\theta}$, and $\hat{\sigma}$ for various number of receiver antennas $M$. Initially, it can be seen in Fig. 3a that as $d_r$ increases, the value of $\hat{k}$ generally decreases, indicating a relatively high probability of small $\phi$ deviations, but also a non-negligible probability of larger $\phi$ deviations. However, for smaller $M$ values, the decrease rate of $\hat{k}$ is smaller, as it is easier to find appropriate RISs that will induce small deviations even with larger dimensions because the $M$ is small. Furthermore, we can see in Fig. 3b and in Fig. 3c that for larger RIS sizes, the values of both $\hat{\theta}$ and $\hat{\sigma}$ increase for all the examined $M$ values, illustrating that if the PWE consists of fewer and larger RISs, the average difference between the desired and actual DoA increases. In addition, considering that an increase of $\theta$ and $\sigma$ leads to wider PDFs for the Gamma and Rayleigh distributions, respectively, it also suggests a greater variance in the observed $\phi$ deviations, since fewer RIS units result in poorer DoA matching resolution. This effect is more severe for larger $M$ values, where the increase rate of both $\hat{\theta}$ and $\hat{\sigma}$ becomes larger, indicating that as the number of antennas increases, the induced deviations for increased RIS size also escalate. In conclusion, Fig. \ref{fig:3} underscores the critical balance between the size and number of RIS units within a PWE and the number of receiver antennas being served.

\begin{figure}
\centering
\begin{subfigure}{.455\textwidth}
 \includegraphics[width=0.99\textwidth]{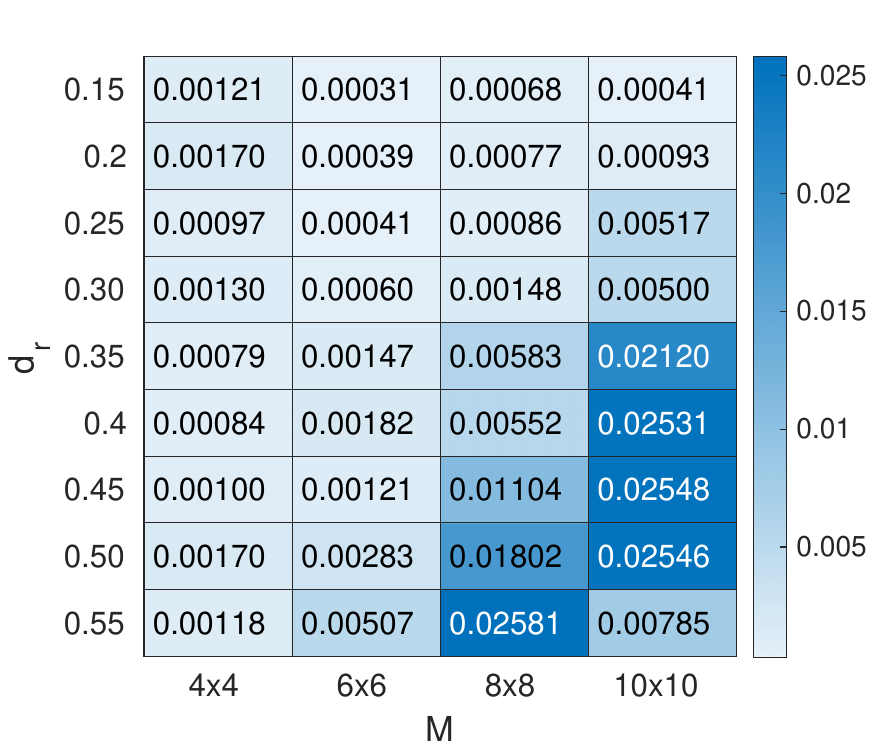}
%    \vspace{3mm}
\end{subfigure}
\begin{subfigure}{.455\textwidth}
\includegraphics[width=0.99\textwidth]{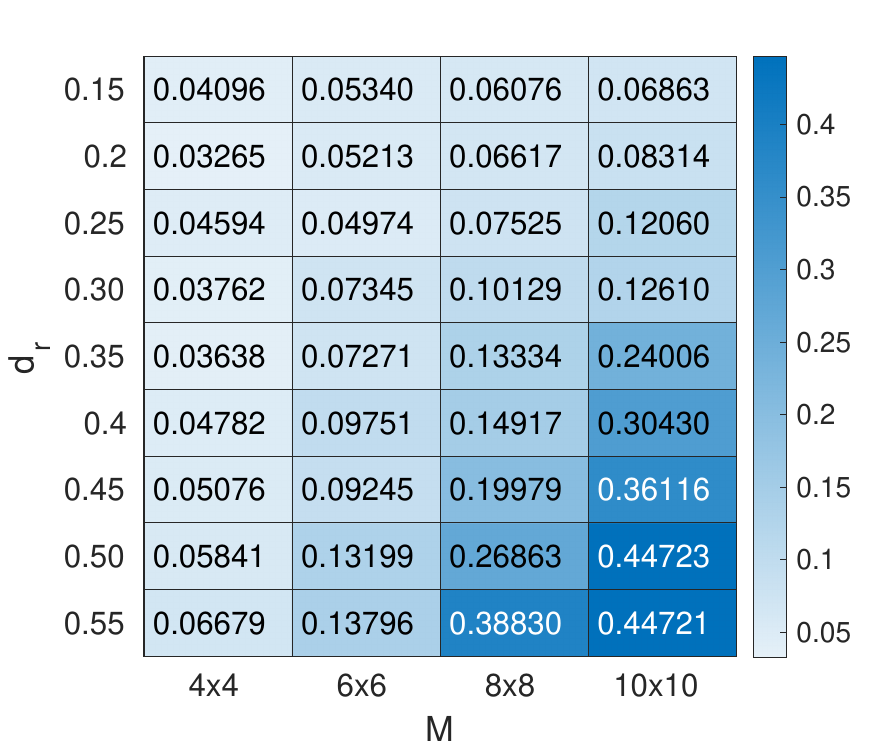}
    \end{subfigure}
\caption{KLD for a) Gamma Distribution b) Rayleigh Distribution}
\label{fig: KL-Div}
\end{figure}

Fig. \ref{fig: KL-Div} shows the KLD between the empirical PDF with the examined distribution models for various RIS sizes and number of receiver antennas. As it can be observed, for smaller RIS dimensions and fewer receiver antennas, both the Gamma and Rayleigh distributions closely approximate the empirical data, indicating a robust model fit under these conditions. Conversely, as $d_r$ and $M$ values increase, the modeling performance degrades. This degradation can be attributed to the increased variance in angle of arrival deviation for larger $M$ and $d_r$ values, a phenomenon that complicates the accurate description of empirical data by the examined distributions. 
Finally, the Gamma distribution consistently outperforms the Rayleigh distribution in data fitting, evidenced by its significantly lower KLD values across all explored combinations of $d_r$ and $M$, underscoring the Gamma distribution's enhanced adaptability in capturing the empirical data's characteristics. In conclusion, while both distributions offer valuable insights, the Gamma distribution's notably lower KLD values across diverse $d_r$ and $M$ settings highlight its efficiency in reflecting the empirical data, marking its significance in the analysis of DoA deviations within a PWE.

\section{Conclusions}\label{conclusion}
In this paper, we examined the novel methodology for RF wavefront copying within PWEs, aimed at advancing XR services through accurate RF wavefront replication at the receiver's antennas. Specifically, we proposed a routing algorithm that minimizes replication errors within the practical constraints of RIS codebooks. Furthermore, our detailed noise modeling and simulation results illustrate how PWE characteristics, such as RIS size and the number of receiver antennas, influence wavefront replication accuracy. Finally, our analysis comparing the Gamma and Rayleigh distribution models with empirical data underlines the Gamma distribution's superior performance, demonstrating its effectiveness in describing the data across different RIS sizes and number of receiver antennas.

\section*{Acknowledgment}
This work has been funded by the FORTH Synergy Grant 2022 `WISAR' and the European Union's Horizon 2020 research and innovation programmes EMERALDS (GA EU101093051) and SENTINEL (GA EU101021659).

\bibliographystyle{IEEEtran}
\bibliography{bibliography}
\end{document}